\documentclass[aip,jcp,reprint,raggedbottom,amsmath,amssymb,superscriptaddress,longbibliography]{revtex4-2}
\usepackage[T1]{fontenc}
\usepackage{bm,graphicx,booktabs}
\usepackage{placeins}
\usepackage{hyperref}
\hypersetup{hidelinks,pdftitle={Electronic excitation spectra and recovery of excited states with neural network wave functions},pdfauthor={Xiaoyu Zhang; Weizhong Fu; Yixiao Chen}}
\newcommand{\ket}[1]{\lvert #1\rangle}
\newcommand{\bra}[1]{\langle #1\rvert}
\newcommand{\braket}[2]{\langle #1\vert #2\rangle}
\newcommand{\dd}{\mathrm{d}}
\newcommand{\ii}{\mathrm{i}}
\newcommand{\R}{\mathbf R}
\newcommand{\Hbar}{\overline H}
\newcommand{\calL}{\mathcal L}
\newcommand{\calF}{\mathcal F}
\newcommand{\E}{\mathbb E}
\newcommand{\Nphi}{N_{\Phi}}
\newcommand{\norm}[1]{\left\lVert #1\right\rVert}
\begin{document}

\title{Electronic excitation spectra and recovery of excited states with neural network wave functions}
\author{Xiaoyu Zhang}
\email{zhangxiaoyu@connect.hku.hk}
\affiliation{ByteDance Seed}
\affiliation{Department of Chemistry, The University of Hong Kong, Hong Kong, China}
\affiliation{College of Chemistry and Molecular Engineering, Peking University, Beijing, China}
\author{Weizhong Fu}
\affiliation{ByteDance Seed}
\affiliation{School of Physics, Peking University, Beijing, China}
\author{Yixiao Chen}
\email{yixiao.chen@bytedance.com}
\affiliation{ByteDance Seed}

\begin{abstract}
Accurate electronic spectra require both a flexible description of electron correlation and a tractable treatment of the many states contributing to the response. We combine neural network wave functions with the Lorentz integral transform to calculate electronic spectra directly in continuous coordinates, without truncation error from a fixed one-electron basis and with polynomial computational cost per optimization step. Instead of constructing a prescribed set of excited states, the method solves an inhomogeneous Schr\"odinger equation at a chosen complex energy. This formulation gives access, in principle, to the entire spectrum coupled to a perturbation, including bound excitations and the ionization continuum, without explicitly determining all lower-lying eigenstates. A finite imaginary energy controls the resolution and keeps the response square integrable. Near an isolated bound excitation, the normalized response also recovers the corresponding eigenstate as the width tends to zero. A helium application illustrates the extraction of an excitation energy and oscillator strength. The formulation provides a route from neural descriptions of electronic correlation to spectra beyond a small manifold of low-lying states.
\end{abstract}

\maketitle

\section{Introduction}

Electronic absorption spectra encode both excitation energies and transition strengths. A broad spectrum requires an accurate description of electron correlation and a tractable treatment of many final states, including the ionization continuum. These are distinct tasks: improving a wave function does not itself provide a spectrum, while an exact spectral calculation in an insufficient orbital space retains its basis error. An effective approach must address both the electronic representation and the extraction of spectral information.

In a finite orbital space, full configuration interaction (FCI) is the exact many-electron reference. Excited-state FCI quantum Monte Carlo (FCIQMC) samples and orthogonalizes evolving states,\cite{Blunt2015} and transition density matrices provide intensities.\cite{Blunt2017} Semistochastic heat-bath configuration interaction instead selects determinants and estimates omitted contributions perturbatively.\cite{Holmes2017} Auxiliary-field QMC also treats orbital Hamiltonians, with practical phaseless and excited-state constraints introducing additional approximations.\cite{Ma2013} Convergence toward FCI tests the correlation treatment for that Hamiltonian; eliminating orbital truncation requires separate basis convergence.

Neural quantum states provide a different compression of the wave function. Their introduction for spin systems\cite{Carleo2017} was followed by neural backflow\cite{Luo2019} and autoregressive sampling.\cite{Sharir2020} In molecular orbital space, restricted Boltzmann machines\cite{Choo2020} and the transformer-based QiankunNet\cite{Shang2025} represent occupation amplitudes. Their exact reference remains FCI in the same orbital and active space: neither neural representation nor exact sampling from the represented distribution removes the Hamiltonian's projection.

Real-space QMC instead evaluates the continuous electronic Hamiltonian with explicitly correlated trial functions.\cite{Foulkes2001} Ceperley and Bernu extended imaginary-time projection to excited-state matrix elements.\cite{Ceperley1988} Subsequent optimization of orbitals and configuration coefficients in Jastrow--Slater functions\cite{Schautz2004} and balanced selected-CI/QMC trial functions\cite{Dash2019} improved molecular excitations. Although Gaussian orbitals may enter these trials, the correlated functions generally extend outside the associated finite FCI space. Appropriate benchmarks are converged results for the continuous Hamiltonian, with fixed-node and pseudopotential approximations identified separately.

Neural networks enlarge this real-space flexibility. PauliNet augments reference orbitals with neural Jastrow and backflow functions,\cite{Hermann2020} whereas FermiNet learns orbitals depending on all electron coordinates without a fixed one-electron expansion.\cite{Pfau2020} Transferable ans\"atze reuse information between molecules,\cite{Scherbela2024} and neural trials improve subsequent fixed-node DMC.\cite{ren_towards_2023} Minimum-step stochastic reconfiguration,\cite{Chen2024} related sample-space updates,\cite{Rende2024} and SPRING\cite{Goldshlager2024} reduce optimization costs. These advances improve the representation of correlation; selecting excitations remains a separate problem.

State selection is not intrinsically neural. The shifted functional of Zhao and Neuscamman targets an eigenstate without first obtaining lower states.\cite{Zhao2016} With finite variational flexibility, however, variance-based optimization can lose the desired root.\cite{Cuzzocrea2020} Energy minimization with overlap penalties suppresses lower states.\cite{Pathak2021} Its application to double excitations\cite{Shepard2022} and subsequent reassessment\cite{Shepard2025} exposed the importance of controlling state mixing. Weighted ensemble objectives impose conditions on both weights and penalties for simultaneous optimization.\cite{Wheeler2024} Formal minima therefore need to be distinguished from the performance of an approximate ansatz and optimizer.

With neural ans\"atze, symmetry and orthogonality enabled lattice excitations,\cite{Choo2018} while molecular calculations used overlap penalties for energies and transition properties\cite{Entwistle2023} and subsequently improved the penalties and spin selection.\cite{Szabo2024} Natural excited-state VMC (NES-VMC) instead minimizes an energy-matrix trace for a determinant of many-electron functions, targeting the lowest $K$ states without overlap penalties.\cite{Pfau2024} Grassmann VMC develops this subspace geometry and reduces parts of the computational overhead through shared features.\cite{Hendry2026} Methods that construct the lowest $K$ states, sequentially or as a subspace, must represent intervening states to reach higher energies in the selected sector. Their cost consequently grows with the state count, although no single scaling exponent applies to every implementation. For orthonormal trial states, the trace equals their summed energy. Starting from the lowest $K$ exact eigenstates, mix only the highest with the next higher eigenstate. The sum increases by $p\Delta$, where $p$ is the added state's probability weight in the normalized mixture and $\Delta>0$ is their energy separation. Small gaps weakly penalize mixing, independently of the representation. This motivates calculating collective spectral weight without separate eigenvectors.

A response calculation does precisely this. Imaginary-time QMC correlations yield spectra through Bayesian inversion.\cite{Caffarel1992} Neural real-time approaches use variational time stepping,\cite{Gutierrez2022} projected evolution,\cite{Sinibaldi2023} refined propagation and error analysis,\cite{Gravina2025} or subspace corrections.\cite{Kahn2026} Such dynamics has reached continuous electronic coordinates.\cite{Nys2024} For two-dimensional spin models, Mendes-Santos et al. obtained spectra by evolving excitations and Fourier transforming their correlations, with the evolution time controlling resolution.\cite{MendesSantos2023} This route avoids stationary-state enumeration but requires accurate propagation.

Frequency methods avoid that propagation. Lanczos DMRG\cite{Hallberg1995} and VMC with projected excitations\cite{Ferrari2018} construct reduced spectral representations; the kernel polynomial method reconstructs spectra from polynomial moments.\cite{Weisse2006} Liu et al. combined it with autoregressive neural wave functions to calculate molecular absorption without explicit excited-state optimization, using STO-3G orbital Hamiltonians and frozen-core active spaces with small-system FCI comparisons.\cite{Liu2025} Their KPNQS method thus addresses spectral extraction, but retains the orbital restriction discussed above.

Complex-energy formulations connect direct response to continuous coordinates. Correction-vector DMRG\cite{Kuhner1999} and dynamical DMRG\cite{Jeckelmann2002} solve inhomogeneous equations at selected frequencies; Hendry and Feiguin implemented neural correction vectors with Monte Carlo estimates.\cite{Hendry2019} The Lorentz integral transform (LIT) replaces continuum scattering states by a square-integrable response at finite imaginary energy.\cite{Efros1994} Its treatment of discrete and continuous contributions is developed in the subsequent review.\cite{Efros2007} Parnes et al. realized continuous-space neural LIT for nuclear photoabsorption using fidelity optimization and complex normalization.\cite{Parnes2026}

Here we describe electronic correlation without a fixed one-electron basis and calculate spectra without constructing individual final eigenstates. The Coulomb Hamiltonian and an independent complex FermiNet provide the response at chosen energy and resolution. We quantify recovery of an isolated bound eigenspace, relate finite resolution to variational error, and derive the solver's operation count. A helium application compares the transition energy with experiment\cite{Eikema1997} and the oscillator strength with high-precision nonrelativistic theory.\cite{Grabowski2011}

\section{Response theory and recovery of excited states}

We use atomic units unless an energy is explicitly given in electronvolts. At fixed nuclear positions, the nonrelativistic electronic Hamiltonian is
\begin{equation}
H=-\frac12\sum_{i=1}^{N_e}\nabla_i^2
-\sum_{iA}\frac{Z_A}{|\mathbf r_i-\mathbf R_A|}
+\sum_{i<j}\frac{1}{|\mathbf r_i-\mathbf r_j|}+E_{\rm nn}.
\label{eq:H}
\end{equation}
Here $N_e$ is the electron number, $\mathbf r_i$ and $\mathbf R_A$ are electron and nuclear coordinates, $Z_A$ is the nuclear charge, and $E_{\rm nn}$ is the constant nuclear repulsion energy. Let $\ket{\Psi_0}$ be a normalized exact ground state, with energy $E_0$, and define $\Hbar=H-E_0$. A Hermitian transition operator $O_a$, whose component is labeled by $a$, produces the centered source
\begin{equation}
\ket{\Phi_a}=\bigl(O_a-\bra{\Psi_0}O_a\ket{\Psi_0}\bigr)\ket{\Psi_0},
\qquad N_{\Phi_a}=\braket{\Phi_a}{\Phi_a}.
\label{eq:source}
\end{equation}
Centering removes the elastic ground-state contribution. We assume $0<N_{\Phi_a}<\infty$ and suppress $a$ temporarily when discussing one source.

To include bound and scattering states without imposing an artificial discretization, let $\mathsf P(\dd\lambda)$ denote the spectral projector of $\Hbar$ in an energy interval $\dd\lambda$. The positive measure
\begin{equation}
\dd\mu(\lambda)=\bra{\Phi}\mathsf P(\dd\lambda)\ket{\Phi}
\label{eq:measure}
\end{equation}
contains all spectral weight accessible from that source. It includes delta-function contributions at discrete transitions and a continuous contribution above ionization thresholds. Its Lorentz transform, at real scan energy $\omega$ and half-width $\eta>0$, is
\begin{equation}
\calL(\omega,\eta)=\int\frac{\dd\mu(\lambda)}{(\lambda-\omega)^2+\eta^2}.
\label{eq:L}
\end{equation}
Define $z=\omega+\ii\eta$ and $A=\Hbar-z$. The response wave function $\ket{X(z)}$ solves
\begin{equation}
A\ket{X(z)}=\ket{\Phi},
\qquad \ket{X(z)}=(\Hbar-z)^{-1}\ket{\Phi}.
\label{eq:inhom}
\end{equation}
Self-adjointness of $\Hbar$ gives $\norm{(\Hbar-z)^{-1}}\leq\eta^{-1}$. Consequently, the solution exists uniquely and is square integrable even when $\omega$ lies in the continuum. Closure then gives the two equivalent exact expressions
\begin{equation}
\calL(\omega,\eta)=\braket{X}{X}
=\frac{1}{\eta}\operatorname{Im}\braket{\Phi}{X}.
\label{eq:Lexact}
\end{equation}
The positive sign of the imaginary part follows from the convention $A=\Hbar-\omega-\ii\eta$.

The normalized Lorentz kernel makes the relation to the spectrum explicit:
\begin{equation}
S_\eta(\omega)=\frac{\eta}{\pi}\calL(\omega,\eta)
=\int\frac{\eta/\pi}{(\lambda-\omega)^2+\eta^2}\,\dd\mu(\lambda).
\label{eq:broadened}
\end{equation}
As $\eta\to0^+$, $S_\eta(\omega)\dd\omega$ tends weakly to $\dd\mu(\omega)$: integration against a smooth test function recovers the corresponding exact spectral integral. Equations~\eqref{eq:inhom}--\eqref{eq:broadened} therefore encompass the entire spectrum coupled to $\Phi$, including its discrete and continuous contributions. At finite $\eta$, each solution gives the response at the selected energy with Lorentzian broadening.

For the response equation, any error $\delta X$ in the domain of $H$ satisfies
\begin{equation}
\begin{aligned}
\norm{A\delta X}^2
&=\norm{(\Hbar-\omega)\delta X}^2+\eta^2\norm{\delta X}^2\\
&\geq\eta^2\norm{\delta X}^2.
\end{aligned}
\label{eq:coercive}
\end{equation}
Equation~\eqref{eq:coercive} bounds the response error by its residual divided by $\eta$. The finite width controls this stability estimate. Accuracy in a neural representation additionally depends on sampling and nonlinear parameter optimization.

Recovery of an excited-state wave function from a correction vector in the zero-width limit was stated in Eq.~(18) of Ref.~\onlinecite{Jeckelmann2002}. We express this limit for a possibly degenerate isolated eigenspace and bound the error at finite width. Let $\omega_n$ be an isolated eigenvalue of $\Hbar$, let $P_n$ project onto its entire eigenspace, and write $Q_n=1-P_n$. Define $\ket{\chi_n}=P_n\ket{\Phi}$ and assume $\norm{\chi_n}>0$. If $\Delta_n>0$ is the distance from $\omega_n$ to the remainder of the spectrum, then
\begin{equation}
\ket{X(\omega_n+\ii\eta)}=\frac{\ii}{\eta}\ket{\chi_n}+\ket{R_n},
\label{eq:pole}
\end{equation}
where
\begin{equation}
\ket{R_n}=Q_n(\Hbar-\omega_n-\ii\eta)^{-1}Q_n\ket{\Phi},
\qquad
\norm{R_n}\leq\frac{\norm{Q_n\Phi}}{\Delta_n}.
\label{eq:remainder}
\end{equation}
The two terms in Eq.~\eqref{eq:pole} are orthogonal. Therefore
\begin{equation}
1-\frac{|\braket{\chi_n}{X}|^2}{\norm{\chi_n}^2\norm{X}^2}
\leq\frac{\eta^2}{\Delta_n^2}
\frac{\norm{Q_n\Phi}^2}{\norm{\chi_n}^2},
\label{eq:polebound}
\end{equation}
and
\begin{equation}
\frac{-\ii\ket{X(\omega_n+\ii\eta)}}{\norm{X(\omega_n+\ii\eta)}}
\longrightarrow\frac{\ket{\chi_n}}{\norm{\chi_n}}
\qquad (\eta\to0^+).
\label{eq:recovery}
\end{equation}
For a nondegenerate level, $\chi_n$ is the eigenfunction multiplied by its transition amplitude, so the recovered normalized state differs from it only by an overall phase. For a degenerate level, one source recovers a particular vector in the degenerate eigenspace, not a complete basis for that space.

Recovery of an individual state consequently requires a width small relative to its separation from the remainder of the spectrum. Weak coupling to the source increases the relative contribution of the remainder in Eq.~\eqref{eq:polebound}. These conditions distinguish an individual bound-state wave function from the combined response at finite resolution.

For later use, the intensity of an isolated line is $I_{an}=\norm{P_n\Phi_a}^2$. Its contribution to the transform can be written
\begin{equation}
\calL_a(\omega,\eta)=\frac{I_{an}}{(\omega_n-\omega)^2+\eta^2}
+B_a(\omega,\eta),
\label{eq:line}
\end{equation}
where $B_a$ contains all remaining bound and continuum weight. Thus an isolated line can yield both its energy and its transition strength from a local transform analysis. For the electronic dipole $\bm\mu=-\sum_i\mathbf r_i$, the isotropic length-gauge oscillator strength, summed over a degenerate final level when necessary, is
\begin{equation}
f_{0n}=\frac{2}{3}\omega_n\sum_{a=x,y,z}I_{an}.
\label{eq:f}
\end{equation}
This is the usual dipole expression with the final-state degeneracy included in $P_n$. The constant nuclear dipole makes no inelastic contribution.

\section{Neural variational formulation}

The implementation first optimizes a real ground-state neural wave function by VMC and then fixes its parameters. In the working expressions below, $\Psi_0$ and $E_0$ denote this variational reference and its estimated energy. The response $X_\theta$ is an independent complex FermiNet, with real trainable parameters $\theta$, real feature layers, and complex orbitals. Unlike the neural Pfaffian used for the nuclear systems of Parnes et al.,\cite{Parnes2026} this representation uses electronic determinants with the Coulomb Hamiltonian of Eq.~\eqref{eq:H}. Its orbitals depend on all electron positions, so the optimized response is not restricted to the Gaussian basis used to initialize the ground state. For helium, with the nucleus at the origin, odd spatial parity is imposed through
\begin{equation}
X_\theta(\R)=\frac{\psi_\theta(\R)-\psi_\theta(-\R)}{2},
\label{eq:parity}
\end{equation}
where $\R=(\mathbf r_1,\ldots,\mathbf r_{N_e})$. This matches the parity of a dipole source from an even ground state. The numbers of spin-up and spin-down electrons are fixed.

To solve the response equation, we align $Y_\theta=A X_\theta$ with the source. Define its fidelity and complex scale by
\begin{equation}
\calF_\theta=
\frac{|\braket{\Phi}{Y_\theta}|^2}
{\Nphi\braket{Y_\theta}{Y_\theta}},
\qquad
c_\theta=\frac{\braket{\Phi}{Y_\theta}}{\Nphi}.
\label{eq:F}
\end{equation}
For nonzero $Y_\theta$, $0\leq\calF_\theta\leq1$, and $\calF_\theta=1$ if and only if $Y_\theta=c_\theta\Phi$. The corrected response is therefore $X_\theta/c_\theta$. In particular, maximizing fidelity alone does not determine the physical amplitude or phase of the uncorrected network output.

With
$\pi_\Phi=|\Phi|^2/\Nphi$ and
$\pi_Y=|Y_\theta|^2/\norm{Y_\theta}^2$, the implemented objective is
\begin{equation}
\mathcal J_\theta=1-\calF_\theta+\lambda\mathcal D_\theta,
\qquad
\mathcal D_\theta=\int\pi_Y\log\frac{\pi_Y}{\pi_\Phi}\,\dd\R.
\label{eq:objective}
\end{equation}
Here $\lambda\geq0$ is the regularization weight, and $\mathcal D_\theta$ is the Kullback--Leibler divergence of $\pi_Y$ from $\pi_\Phi$. When finite, this term penalizes mismatched probability distributions and vanishes with the fidelity loss at an exact solution. 

The source is independent of the scan frequency, allowing its configurations to be reused. In contrast to direct sampling from $\pi_\Phi$ in Ref.~\onlinecite{Parnes2026}, the electronic implementation adds a floor to the magnitude of the dipole factor used for sampling. For a dipole component, write $d_a(\R)=\mu_a(\R)-\langle\mu_a\rangle_0$ and $\Phi_a=d_a\Psi_0$. The sampling distribution and its correction weights are
\begin{equation}
\begin{aligned}
b_a(\R)&=\max\bigl(|d_a(\R)|,\nu\bigr),\\
q_\nu(\R)&\propto |\Psi_0(\R)|^2b_a(\R)^2,
&v_a(\R)&=\frac{d_a(\R)^2}{b_a(\R)^2},
\end{aligned}
\label{eq:q}
\end{equation}
where $\nu>0$ is a sampling floor. Expectations over the physical normalized source are obtained from
\begin{equation}
\E_{\pi_\Phi}[g]=\frac{\E_{q_\nu}[v_a g]}{\E_{q_\nu}[v_a]}.
\label{eq:weights}
\end{equation}
The floor changes the sampling distribution, while Eq.~\eqref{eq:weights} retains the unmodified dipole source in all physical expectations.

Let $u_\theta(\R)=Y_\theta(\R)/\Phi(\R)$ and $t_\theta(\R)=X_\theta(\R)/\Phi(\R)$, defined off the source nodes. Introduce the moments
\begin{equation}
C_\theta=\E_{\pi_\Phi}[u_\theta],\qquad
B_\theta=\E_{\pi_\Phi}[|u_\theta|^2],\qquad
T_\theta=\E_{\pi_\Phi}[t_\theta].
\label{eq:moments}
\end{equation}
The working objective and transform are then
\begin{equation}
\calF_\theta=\frac{|C_\theta|^2}{B_\theta},\qquad
\mathcal D_\theta=
\frac{\E_{\pi_\Phi}[|u_\theta|^2\log|u_\theta|^2]}{B_\theta}
-\log B_\theta,
\label{eq:mcF}
\end{equation}
\begin{equation}
\calL_\theta=\frac{\Nphi}{\eta}\operatorname{Im}\frac{T_\theta}{C_\theta}
=\frac{1}{\eta}\operatorname{Im}
\frac{\braket{\Phi}{X_\theta}}{c_\theta}.
\label{eq:mcL}
\end{equation}
The source norm is estimated with sampling from the ground state. Equation~\eqref{eq:weights} is applied to the moments evaluated from the source configurations in Eqs.~\eqref{eq:moments}--\eqref{eq:mcF}. The ratios of sample averages are consistent estimators of these quantities. The computed transform is the signed overlap in Eq.~\eqref{eq:mcL}; it equals the response norm in Eq.~\eqref{eq:Lexact} when the response equation is satisfied.

A direct bound connects the fidelity to error in solving the response equation. For $c_\theta\ne0$, let
$r_\theta=A(X_\theta/c_\theta)-\Phi$. From Eq.~\eqref{eq:F},
\begin{equation}
\frac{\norm{r_\theta}}{\norm{\Phi}}
=\sqrt{\frac{1-\calF_\theta}{\calF_\theta}}.
\label{eq:residual}
\end{equation}
Applying the resolvent bound to this residual gives
\begin{equation}
\norm{X_\theta/c_\theta-X}
\leq\frac{\sqrt{\Nphi}}{\eta}
\sqrt{\frac{1-\calF_\theta}{\calF_\theta}},
\label{eq:solutionbound}
\end{equation}
and the overlap estimate satisfies
\begin{equation}
|\calL_\theta-\calL|
\leq\frac{\Nphi}{\eta^2}
\sqrt{\frac{1-\calF_\theta}{\calF_\theta}}.
\label{eq:Lbound}
\end{equation}
These conservative bounds hold for the fixed source and reference energy whenever $\calF_\theta>0$. Relative to the projected error estimates in Eqs.~(6)--(7) of Ref.~\onlinecite{Parnes2026}, the bound in Eq.~\eqref{eq:Lbound} uses the uniform resolvent norm in place of factors depending on the response and does not require an expansion in small infidelity. The explicit width dependence shows how the required accuracy tightens as the resolution improves.

The electronic Hamiltonian acts on the complex response through spatial automatic differentiation. If $h_\theta=\log X_\theta$ locally away from nodes and $V$ is the potential in Eq.~\eqref{eq:H}, the local energy is
\begin{equation}
\frac{H X_\theta}{X_\theta}
=-\frac12\sum_i\left[\nabla_i^2h_\theta+
\nabla_i h_\theta\mathbin{\cdot}\nabla_i h_\theta\right]+V.
\label{eq:complexkinetic}
\end{equation}
The gradient product has no complex conjugation. The implementation obtains the full spatial Hessians of the real and imaginary parts of $h_\theta$ and differentiates the resulting Hamiltonian action with respect to $\theta$.

To specify the parameter update, let $g_\theta=\nabla_\theta\calF_\theta-\lambda\nabla_\theta\mathcal D_\theta$ and define the centered complex logarithmic derivatives
\begin{equation}
\Delta O_\alpha(\R)=\partial_{\theta_\alpha}\log Y_\theta(\R)
-\E_{\pi_Y}[\partial_{\theta_\alpha}\log Y_\theta].
\label{eq:score}
\end{equation}
The real metric and damped update are
\begin{equation}
S_{\alpha\beta}=\operatorname{Re}\E_{\pi_Y}
[\Delta O_\alpha^*\Delta O_\beta],
\label{eq:metric}
\end{equation}
\begin{equation}
(S+\epsilon I)d=g_\theta+\epsilon\mu d_{\rm prev}.
\label{eq:update}
\end{equation}
The damping is proportional to $\operatorname{tr}S/P$, subject to a positive lower bound, where $P$ is the number of real parameters. The previous unscaled direction is retained as $d_{\rm prev}$, and the applied step is
\begin{equation}
\Delta\theta=\min\!\left(\gamma,\frac{d_{\max}}{\norm{d}}\right)d,
\label{eq:step}
\end{equation}
with learning rate $\gamma$ and maximum step norm $d_{\max}$. This damped natural gradient update retains the preceding direction in the manner of SPRING.\cite{Goldshlager2024} Equations~\eqref{eq:score}--\eqref{eq:step} specify the metric and normalization used in the electronic implementation.

\section{Computational scaling}

Sample-space solves underlie minimum-step stochastic reconfiguration,\cite{Chen2024} while a related linear algebra formulation connects parameter-space and sample-space updates.\cite{Rende2024} We count both forms below for the present implementation, including the spatial derivatives required by the response objective.

The operation count has two contributions: evaluation and differentiation of the electronic wave function, and solution of the metric equation. Let $N_A$ be the number of nuclei, $D$ the number of determinants, $L$ the number of feature layers, and $h$ an upper bound on their widths. Evaluation of the pair features costs $O(LN_e^2h^2)$; construction of the orbital matrices and their nuclear envelopes costs $O(DN_e^2h+DN_e^2N_A)$; and the $D$ determinant factorizations cost $O(DN_e^3)$. Including the initial electron--nucleus features gives the bound
\begin{equation}
\begin{aligned}
C_\psi=O\bigl(&LN_e^2h^2+N_eN_Ah\\
&+DN_e^2h+DN_e^2N_A+DN_e^3\bigr).
\end{aligned}
\label{eq:evalcost}
\end{equation}
For fixed $D$, $L$, and $h$, and $N_A=O(N_e)$, this is cubic in $N_e$. Complex orbitals and the parity projection change the prefactor rather than the power.

The implemented Hessian has $3N_e$ coordinate directions. Forward differentiation of a reverse-mode gradient therefore costs $O(N_eC_\psi)$ for the Hamiltonian action. Reverse differentiation with respect to the parameters gives its logarithmic derivatives with the same arithmetic order, together with the cost of writing the $P$ derivatives. For a batch of $B$ configurations,
\begin{equation}
C_{\rm der}=O\!\left[B(N_eC_\psi+P)\right].
\label{eq:dercost}
\end{equation}
Thus the electronic differentiation contributes $O(BDN_e^4)$ at fixed feature dimensions. Evaluation of the Coulomb potential, which costs $O(N_e^2+N_eN_A+N_A^2)$, is lower order under the same assumptions. The count includes the mixed spatial and parameter derivatives; it is not based solely on evaluating the neural network.

To count the optimization cost, stack the weighted real and imaginary parts of the centered derivatives in Eq.~\eqref{eq:score} into a real matrix $J$ with $m=2B$ rows and $P$ columns, so that the sampled metric is $S=J^{\mathsf T}J$. When $P\leq m$, the implementation forms and factorizes the $P\times P$ matrix. When $P>m$, it uses the equivalent $m\times m$ system. For $b=g_\theta+\epsilon\mu d_{\rm prev}$, the latter solution is
\begin{equation}
 d=\frac{1}{\epsilon}\left[b-J^{\mathsf T}
 (JJ^{\mathsf T}+\epsilon I_m)^{-1}Jb\right].
\label{eq:dual}
\end{equation}
The dense matrix multiplication and Cholesky factorization therefore cost
\begin{equation}
 C_{\rm metric}=\begin{cases}
 O(mP^2+P^3),&P\leq m,\\
 O(m^2P+m^3),&P>m.
 \end{cases}
\label{eq:metriccost}
\end{equation}
Storage of the derivatives and the smaller metric matrix requires $O(mP+\min(m,P)^2)$ numbers, in addition to the neural differentiation workspace and electronic configurations. The complete optimization step has the polynomial cost
\begin{equation}
 C_{\rm step}=O\!\left[B(N_eC_\psi+P)\right]+C_{\rm metric}.
\label{eq:stepcost}
\end{equation}
The shared feature transformations, orbital outputs, and nuclear envelopes give $P=O(Lh^2+N_Ah+DN_eh+DN_eN_A)$. Thus $P=O(N_e^2)$ at fixed $D$, $L$, and $h$ with $N_A=O(N_e)$. In the asymptotic dual branch, a fixed batch size then leaves the $O(N_e^4)$ electronic differentiation as the leading operation count per step.

For $N_\omega$ frequencies with $T_j$ updates at frequency $j$, the response optimization costs $\sum_{j=1}^{N_\omega}T_jC_{{\rm step},j}$, in addition to preparation of the ground state and sampling pools. Generating $M$ configurations with $n_{\rm MC}$ Metropolis moves per configuration costs $O(Mn_{\rm MC}C_{\psi_0})$; these source configurations can be reused across frequencies. Polynomial scaling here refers to arithmetic at specified architecture, sample size, and optimization effort. The number of samples and iterations required for a fixed physical accuracy remains dependent on the system and spectral resolution.

\section{Results and discussion}

We apply the method to the lowest dipole-allowed excitation of helium. Both the ground-state and response networks contain 16 full determinants and four feature layers, with one-electron and two-electron widths of 64 and 16, respectively. Ground-state orbital pretraining uses an aug-cc-pVTZ reference for 2000 iterations, followed by 50\,000 VMC iterations. The Gaussian basis is used only for initialization. The subsequent optimization and response calculation take place directly in electronic coordinates.

\begin{figure*}[t]
\centering
\includegraphics[width=0.80\textwidth]{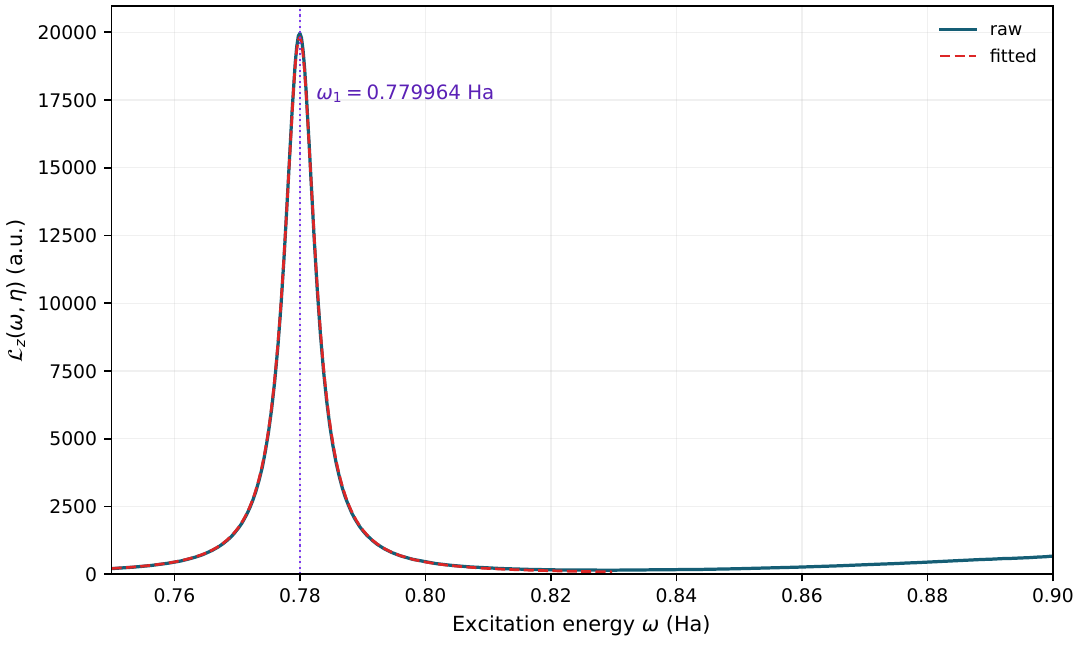}
\caption{Lorentz transform of the $z$-polarized dipole response of helium at $\eta=0.003$ hartree. The solid curve is the calculated transform. The dashed curve is the fit of Eq.~\eqref{eq:fit}, determined in the interval 0.750--0.830 hartree and continued for display. The vertical dotted line marks the fitted lowest dipole-allowed transition.}
\label{fig:he}
\end{figure*}

\begin{table*}[t]
\caption{Excitation energies and oscillator strengths for the lowest dipole-allowed helium line. Energy deviations are relative to the experimental value of Eikema et al.;\cite{Eikema1997} strength deviations are relative to the infinite-nuclear-mass, nonrelativistic calculation of Grabowski and Chernoff.\cite{Grabowski2011}}
\label{tab:he}
\centering
\renewcommand{\arraystretch}{1.1}
\begin{ruledtabular}
\begin{tabular}{lrrrr}
Method / basis & Energy (eV) & Energy deviation (eV) & Oscillator strength & Strength deviation (\%)\\
FCI / cc-pVTZ\textsuperscript{a} &51.169214&$+29.951191$&2.251926&$+715.4$\\
FCI / aug-cc-pVDZ\textsuperscript{a}&27.367841&$+6.149818$&1.312244&$+375.2$\\
FCI / aug-cc-pVTZ\textsuperscript{a}&25.361727&$+4.143704$&1.065781&$+285.9$\\
FCI / aug-cc-pVQZ&24.041187&$+2.823164$&0.882799&$+219.7$\\
FCI / d-aug-cc-pVDZ&21.139482&$-0.078541$&0.314200&$+13.8$\\
FCI / d-aug-cc-pVTZ&21.324760&$+0.106737$&0.347586&$+25.9$\\
FCI / d-aug-cc-pVQZ&21.299371&$+0.081348$&0.329295&$+19.2$\\
NQS--LIT&21.223892&$+0.005869$&0.278055&$+0.68$\\
Experiment&21.218023&---&---&---\\
Nonrelativistic theory&---&---&0.27616499(27)&---\\
\end{tabular}
\end{ruledtabular}
\vspace{3pt}
\parbox{\textwidth}{\footnotesize\textsuperscript{a}These finite-basis FCI roots lie above the He$^++e^-$ ionization threshold and do not provide converged bound-state transition energies. FCI energies and transition dipole moments were obtained with PySCF.\cite{10.1063/5.0337441}}
\end{table*}

The source is the $z$ component of the dipole, centered at zero by atomic symmetry, with sampling floor $\nu=10^{-4}$ in atomic units. The training and evaluation pools are generated separately, with 20 Metropolis steps per update. The response objective uses $\lambda=1$, learning rate $\gamma=0.002$, decay $\mu=0.99$, and maximum step norm $d_{\max}=0.02$. Up to 6000 iterations are allowed at each frequency; the checkpoint is selected using the evaluation pool and a plateau stopping criterion.

The scan uses $\eta=0.003$ hartree and 601 requested frequencies between 0.750 and 0.900 hartree. Parameters from the preceding accepted solution initialize the next frequency, with additional continuation points inserted when needed. To extract the lowest line, we fit the interval 0.750--0.830 hartree to
\begin{equation}
\calL_z^{\rm fit}(\omega)=
\frac{I_z}{(\omega_1-\omega)^2+\eta^2}+b_0,
\label{eq:fit}
\end{equation}
using unweighted least squares, with $I_z\geq0$ and $\omega_1$ restricted to this interval. The constant $b_0$ approximates the slowly varying contribution of other transitions within the fitting window. This local extraction of a bound line differs from the continuum inversion used in the nuclear applications of Ref.~\onlinecite{Parnes2026}. For the isotropic helium ground state, the three polarization components have equal total intensity, so Eq.~\eqref{eq:f} yields the oscillator strength of the complete atomic line from one component:
\begin{equation}
f_{01}=2\omega_1I_z.
\label{eq:hefactor}
\end{equation}

Figure~\ref{fig:he} shows the calculated transform and its local fit. The resulting excitation energy is 21.223892 eV and the oscillator strength is 0.278055. The physical line is conventionally assigned as $1s^2\,{}^1S_0\rightarrow1s2p\,{}^1P_1^{\circ}$; the electronic calculation uses the nonrelativistic $1\,{}^1S\rightarrow2\,{}^1P$ transition. Eikema et al. measured the $^4$He transition frequency as $5\,130\,495\,083(45)$ MHz,\cite{Eikema1997} corresponding to $21.21802278(19)$ eV, or 21.218023 eV at the precision of Table~\ref{tab:he}. The energy deviation is therefore $+0.005869$ eV relative to experiment.

For the oscillator strength, Grabowski and Chernoff obtain $f=0.27616499(27)$ by a pseudospectral solution of the two-electron Schr\"odinger equation.\cite{Grabowski2011} Their Hamiltonian explicitly uses infinite nuclear mass and the nonrelativistic approximation [Eqs.~(2)--(3) of that work], matching Eq.~\eqref{eq:H} for helium. Their result in Sec.~VII~A provides a theoretical reference for the same Hamiltonian, relative to which our oscillator strength differs by $+0.68\%$. The energy comparison instead uses the measured physical $^4$He line, which also contains finite-mass and relativistic contributions.

The full configuration interaction (FCI) results in Table~\ref{tab:he} illustrate the sensitivity of this diffuse excitation to the orbital space. Adding diffuse functions changes both the excitation energy and the oscillator strength substantially. Double augmentation improves the transition energy, but the excitation energies are not monotonic because both the ground and excited total energies change with the basis. The intensities remain appreciably more sensitive. The neural calculation is closer to the reference values than the listed FCI calculations and avoids their fixed one-electron basis restriction. Its remaining numerical accuracy is governed by the neural representation, optimization, and sampling rather than by a truncated orbital expansion.

At higher frequencies the scan becomes more difficult to converge. A plausible explanation is the spatial shape of the sampling distribution. The distribution $q_\nu$ in Eq.~\eqref{eq:q} inherits the compact decay of the ground-state density, multiplied by the dipole factor, whereas higher excited responses can extend farther from the nucleus and contain additional radial structure. A finite pool drawn from $q_\nu$ may then sample poorly the regions important for the response and its parameter derivatives. The floor regularizes the dipole nodes but does not substantially broaden the asymptotic tail of the proposal. This is a possible limitation during optimization, not a mismatch required at the exact solution, for which $Y_\theta/c_\theta=\Phi$. The importance of covering both the source and the Hamiltonian action during optimization was already emphasized by Hendry and Feiguin.\cite{Hendry2019} More broadly distributed or adaptive configurations may improve convergence here.

\FloatBarrier
\section{Conclusions}

Electronic neural wave functions combined with the LIT give access, in principle, to the entire spectrum coupled to a perturbation without a fixed one-electron basis. The target is a response at specified energy and resolution. Finite width treats bound and continuum contributions within one square-integrable function; the limit at an isolated pole recovers the bound eigenstate. For fixed architecture hyperparameters, sampling effort, frequency grid, and iteration counts, the overall optimization cost scales polynomially with electron number.

The helium application demonstrates extraction of a transition energy and oscillator strength, while reliable higher excitations require further algorithmic development. The compact source distribution may inadequately sample more extended and structured responses, suggesting that broader or adaptive sampling should be investigated together with control of the response residual.

\begin{acknowledgments}
We thank Ruichen Li for carefully reviewing the code implementation associated with one of our previous theoretical works. We thank the ByteDance Seed AI for Science teams for their invaluable support.
\end{acknowledgments}

\section*{Author contributions}
Xiaoyu Zhang and Yixiao Chen conceived and initiated the project and developed preliminary approaches. Xiaoyu Zhang independently developed the final theoretical framework and wrote the manuscript. Weizhong Fu contributed through project-related discussions. Xiaoyu Zhang is the sole first author; Xiaoyu Zhang and Yixiao Chen are co-corresponding authors.

\section*{Data availability}
The numerical results reported in this work are included in the article. The computational input files and execution and postprocessing scripts are provided with the public implementation identified below.

\section*{Code availability}
{\raggedright
The implementation is publicly available in the JaQMC repository\cite{jaqmc} at
\url{https://github.com/bytedance/jaqmc/tree/main/contrib/lit}.
The example directory \nolinkurl{contrib/lit/examples/he} contains the helium inputs and scripts. The LIT contribution is identified by commit
\href{https://github.com/bytedance/jaqmc/commit/5234d747d690dd8e0d09bb7024b3deb6f263b3dd}{\texttt{5234d747}} (September 4, 2026).\par}

\makeatletter\immediate\write\@auxout{\string\citation{AIPControl}}\makeatother
\bibliographystyle{aipnum4-2}
\bibliography{main}
\end{document}